\documentclass[11pt]{article}
\usepackage{amssymb,amsmath,amsfonts,epsfig}
\newcommand{\sect}[1]{\setcounter{equation}{0}\section{#1}}

\begin{document}
\setlength{\baselineskip}{5.0mm}
\title{Symmetry Classes}
\thispagestyle{empty}
\maketitle
\begin{center}
{{\sc Martin R.\ Zirnbauer}\\~\\ Institut f\"ur
Theoretische Physik, Universit\"at zu K\"oln,\newline Z\"ulpicher
Stra{\ss}e 77, 50937 K\"oln, Germany}

\bigskip
{\bf Abstract}
\end{center}
Physical systems exhibiting stochastic or chaotic behavior are often
amenable to treatment by random matrix models. In deciding on a good
choice of model, random matrix physics is constrained and guided by
symmetry considerations. The notion of `symmetry class' (not to be
confused with `universality class') expresses the relevance of
symmetries as an organizational principle. Dyson, in his 1962 paper
referred to as The Threefold Way, gave the prime classification of
random matrix ensembles based on a quantum mechanical setting with
symmetries. In this article we review Dyson's Threefold Way from a
modern perspective. We then describe a minimal extension of Dyson's
setting to incorporate the physics of chiral Dirac fermions and
disordered superconductors. In this minimally extended setting, where
Hilbert space is replaced by Fock space equipped with the
anti-unitary operation of particle-hole conjugation, symmetry classes
are in one-to-one correspondence with the large families of
Riemannian symmetric spaces.

\sect{Introduction}\label{intro}

In Chapter 2 of this handbook\footnote{The present article is to be
Chapter 3 of the Oxford Handbook of Random Matrix Theory.}, the
historical narrative by Bohigas and Weidenm\"uller describes how
random matrix models emerged from quantum physics, more precisely
from a statistical approach to the strongly interacting many-body
system of the atomic nucleus. Although random matrix theory is
nowadays understood to be of relevance to numerous areas of physics,
mathematics, and beyond, quantum mechanics is still where many of its
applications lie. Quantum mechanics also provides a natural framework
in which to classify random matrix ensembles.

In this thrust of development, a symmetry classification of random
matrix ens\-embles was put forth by Dyson in his 1962 paper \emph{The
Threefold Way: alge\-braic structure of symmetry groups and ensembles
in quantum mechanics}, where he proved (quote from the abstract of
\cite{dyson}) ``that the most general matrix ensemble, defined with a
symmetry group which may be completely arbitrary, reduces to a direct
product of independent irreducible ensembles each of which belongs to
one of the three known types''. The three types known to Dyson were
ensembles of matrices which are either complex Hermitian, or real
symmetric, or quaternion self-dual. It is widely acknowledged that
Dyson's Threefold Way has become fundamental to various areas of
theoretical physics, including the statistical theory of complex
many-body systems, mesoscopic physics, disordered electron systems,
and the field of quantum chaos.

Over the last decade, a number of random matrix ensembles beyond
Dyson's classification have come to the fore in physics and
mathematics. On the physics side these emerged from work \cite{Ver94}
on the low-energy Dirac spectrum of quantum chromodynamics, and also
from the mesoscopic physics of low-energy quasi-particles in
disordered superconductors \cite{AZ}. In the mathematical research
area of number theory, the study of statistical correlations of the
values of Riemann zeta and related $L$-functions has prompted some of
the same generalizations \cite{KS}. It was observed early on
\cite{AZ} that these post-Dyson ensembles, or rather the underlying
symmetry classes, are in one-to-one correspondence with the large
families of symmetric spaces.

The prime emphasis of the present handbook article will be on
describing Dyson's fundamental result from a modern perspective. A
second goal will be to introduce the post-Dyson ensembles. While
there seems to exist no unanimous view on how these fit into a
systematic picture, here we will follow \cite{HHZ} to demonstrate
that they emerge from Dyson's setting upon replacing the plain
structure of Hilbert space by the more refined structure of Fock
space.\footnote{We mention in passing that a classification of Dirac
Hamiltonians in two dimensions has been proposed in \cite{BL02}.
Unlike ours, this is not a symmetry classification in Dyson's sense.}
The reader is advised that some aspects of this story are treated in
a more leisurely manner in the author's encyclopedia article
\cite{EMP}.

To preclude any misunderstanding, let us issue a clarification of
language right here: `symmetry class' must not be confused with
`universality class'! Indeed, inside a symmetry class as understood
in this article various types of physical behavior are possible. (For
example, random matrix models for weakly disordered time-reversal
invariant metals belong to the so-called Wigner-Dyson symmetry class
of real symmetric matrices, and so do Anderson tight-binding models
with real hopping and strong disorder. The former are believed to
exhibit the universal energy level statistics given by the Gaussian
Orthogonal Ensemble, whereas the latter have localized eigenfunctions
and hence level statistics which is expected to approach the Poisson
limit when the system size goes to infinity.) For this reason the
present article must refrain from writing down explicit formulas for
joint eigenvalue distributions, which are available only in certain
universal limits.

\sect{Dyson's Threefold Way}\label{sect:dyson}

Dyson's classification is formulated in a general and simple
mathematical setting which we now describe. First of all, the
framework of quantum theory calls for the basic structure of a
complex vector space $V$ carrying a Hermitian scalar product $\langle
\cdot , \cdot \rangle : \, V \times V \to \mathbb{C}\,$. (Dyson
actually argues \cite{dyson} in favor of working over the real
numbers, but we will not follow suit in this respect.) For technical
simplicity, we do join Dyson in taking $V$ to be finite-dimensional.
In applications, $V \simeq \mathbb{C}^n$ will usually be the
truncated Hilbert space of a family of disordered or quantum chaotic
Hamiltonian systems.

The Hermitian structure of the vector space $V$ determines a group
$\mathrm{U}(V)$ of unitary transformations of $V$. Let us recall that
the elements $g \in \mathrm{U}(V)$ are $\mathbb{C}$-linear operators
satisfying the condition $\langle g v , g v^\prime \rangle = \langle
v , v^\prime \rangle$ for all $v,v^\prime \in V$.

Building on the Hermitian vector space $V$, Dyson's setting
stipulates that $V$ be equipped with a unitary group action
\begin{equation}
    G_0 \times V \to V , \quad (g,v) \mapsto \rho_V(g) v \;,
    \quad \rho_V(g) \in \mathrm{U}(V) \;.
\end{equation}
In other words, there is some group $G_0$ whose elements $g$ are
represented on $V$ by unitary operators $\rho_V(g)$. This group $G_0$
is meant to be the group of joint (unitary) symmetries of a family of
quantum mechanical Hamiltonian systems with Hilbert space $V$. We
will write $\rho_V(g) \equiv g$ for short.

Now, not every symmetry of a quantum system is of the canonical
unitary kind. The prime counterexample is the operation, $T$, of
inverting the time direction, called time reversal for short. It is
represented on Hilbert space $V$ by an \emph{anti}-unitary operator
$T \equiv \rho_V(T)$, which is to say that $T$ is complex anti-linear
and preserves the Hermitian scalar product up to complex conjugation:
\begin{equation}
    T (z v) = \overline{z} \, T v , \qquad
    \left\langle T v , T v^\prime \right\rangle
    = \overline{\left\langle v , v^\prime \right\rangle }
    \qquad (z \in {\mathbb C}\,; \; v, v^\prime \in V) \;.
\end{equation}
Another operation of this kind is charge conjugation in relativistic
theories such as the Dirac equation for the electron and its
anti-particle, the positron.

Thus in Dyson's general setting one has a so-called symmetry group $G
= G_0 \cup G_1$ where the subgroup $G_0$ is represented on $V$ by
unitaries, while $G_1$ (not a group) is represented by
anti-unitaries. By the definition of what is meant by a `symmetry',
the generator of time evolution, the Hamiltonian $H$, of the quantum
system is fixed by conjugation $gHg^{-1} = H$ with any $g \in G$.

The set $G_1$ may be empty. When it is not, the composition of any
two elements of $G_1$ is unitary, so every $g \in G_1$ can be
obtained from a fixed element of $G_1\,$, say $T$, by right
multiplication with some $U \in G_0\,$: $g = T U$. The same goes for
left multiplication, i.e., for every $g \in G_1$ there also exists
$U^\prime \in G_0$ so that $g = U^\prime T$. In other words, when
$G_1$ is non-empty, $G_0 \subset G$ is a proper normal subgroup and
the factor group $G / G_0 \simeq \mathbb{Z}_2$ consists of exactly
two elements, $G_0$ and $T G_0 = G_1\,$. For future use we record
that conjugation $U \mapsto T U T^{-1} =: a(U)$ by time reversal is
an automorphism of $G_0\,$.

Following Dyson \cite{dyson} we assume that the special element $T$
represents an {\it inversion} symmetry such as time reversal or
charge conjugation. $T$ must then be a (projective) involution, i.e.,
$T^2 = z \times {\rm Id}_V$ with $0 \not= z \in \mathbb{C}\,$, so
that conjugation by $T^2$ is the identity operation. Since $T$ is
anti-unitary, $z$ must have modulus $|z| = 1\,$, and by the
$\mathbb{C}$-antilinearity of $T$ the associative law
\begin{equation}\label{eq:Tass}
    z\, T = T^2 \cdot T = T \cdot T^2 = T z = \overline{z}\, T
\end{equation}
forces $z$ to be real, which leaves only two possibilities: $T^2 =
\pm \mathrm{Id}_V\,$.

Let us record here a concrete example of some historical importance:
the Hilbert space $V$ might be the space of totally anti-symmetric
wave functions of $n$ particles distributed over the shell-model
space of an atom or an atomic nucleus, and the symmetry group $G$
might be $G = \mathrm{O}_3 \cup T \mathrm{O}_3\,$, the full rotation
group $\mathrm{O}_3$ (including parity) together with its translate
by time reversal $T$.

In summary, Dyson's setting assumes two pieces of data:
\begin{itemize}
{\item[$\bullet$] a finite-dimensional complex vector space $V$
with Hermitian structure,}
{\item[$\bullet$] a group $G = G_0 \cup T G_0$ acting on $V$ by
unitary and anti-unitary operators.}
\end{itemize}
It should be stressed that, in principle, the primary object is the
Hamiltonian, and the symmetries $G$ are secondary objects derived
from it. However, adopting Dyson's standpoint we now turn tables to
view the symmetries as fundamental and given and the Hamiltonians as
derived objects. Thus, fixing any pair $(V,G)$ our goal is to
elucidate the structure of the space of all \emph{compatible}
Hamiltonians, i.e., the self-adjoint operators $H$ on $V$ which
commute with the $G$-action. Such a space is reducible in general:
the $G$-compatible Hamiltonian matrices decompose as a direct sum of
blocks. The goal of classification is to enumerate the irreducible
blocks that occur in this setting.

While the main objects to classify are the spaces of compatible
Hamiltonians $H$, we find it technically convenient to perform some
of the discussion at the integrated level of {\it time evolutions}
$U_t = {\rm e}^{-{\rm i}t H /\hbar}$ instead. This change of focus
results in no loss, as the Hamiltonians can always be retrieved by
linearization in $t$ at $t = 0\,$. The compatibility conditions for
$U \equiv U_t$ read
\begin{equation}\label{eq:G-comp}
    U = g_0^{\vphantom{-1}} U g_0^{-1} = g_1^{\vphantom{-1}} U^{-1}
    g_1^{-1} \quad (\mbox{for all } g_\sigma \in G_\sigma ) \;.
\end{equation}

The strategy will be to make a reduction to the case of the trivial
group $G_0 = \{ \mathrm{Id} \}$. The situation with trivial $G_0$ can
then be handled by enumeration of a finite number of possibilities.

\subsection{Reduction to the case of $G_0 = \{ \mathrm{Id} \}$}
\label{sect:reduce}

To motivate the technical reduction procedure below, we begin by
elaborating the example of the rotation group $\mathrm{O}_3$ acting
on a Hilbert space of shell-model states. Any Hamiltonian which
commutes with $G_0 = \mathrm{O}_3$ conserves total angular momentum,
$L$, and parity, $\pi$, which means that all Hamiltonian matrix
elements connecting states in sectors of different quantum numbers
$(L,\pi)$ vanish identically. Thus, the matrix of the Hamiltonian
with respect to a basis of states with definite values of $(L,\pi)$
has diagonal block structure. $\mathrm{O}_3 $-symmetry further
implies that the Hamiltonian matrix is diagonal with respect to the
orthogonal projection, $M$, of total angular momentum on some axis in
position space. Moreover, for a suitable choice of basis the matrix
will be the same for each $M$-value of a given sector $(L,\pi)$.

To put these words into formulas, we employ the mathematical notions
of orthogonal sum and tensor product to decompose the shell-model
space as
\begin{equation}
    V \simeq \bigoplus_{L \ge 0; \; \pi = \pm 1} V_{(L,\pi)}
    \;, \qquad V_{(L,\pi)} = \mathbb{C}^{m(L,\pi)} \otimes
    \mathbb{C}^{2L+1} \;,
\end{equation}
where $m(L,\pi)$ is the multiplicity in $V$ of the
$\mathrm{O}_3$-representation with quantum numbers $(L,\pi)$. The
statement above is that all symmetry operators and compatible
Hamiltonians are diagonal with respect to this direct sum, and within
a fixed block $V_{(L,\pi)}$ the Hamiltonians act on the first factor
$\mathbb{C}^{m(L,\pi)}$ and are trivial on the second factor
$\mathbb{C}^{2L+1}$ of the tensor product, while the symmetry
operators act on the second factor and are trivial on the first
factor. Thus we may picture each sector $V_{(L,\pi)}$ as a
rectangular array of states where the Hamiltonians act, say,
horizontally and are the same in each row of the array, while the
symmetries act vertically and are the same in each column.

This concludes our example, and we now move on to the general case of
any group $G_0$ acting reductively on $V$. To handle it, we need some
language and notation as follows. A $G_0$-representation $X$ is a
$\mathbb{C}$-vector space carrying a $G_0$-action $G_0 \times X \to
X$ by $(g,x) \mapsto \rho_X(g) x$. If $X$ and $Y$ are
$G_0$-representations, then by the space $\mathrm{Hom}_{G_0}(X,Y)$ of
$G_0$-equivariant homomorphisms from $X$ to $Y$ one means the complex
vector space of $\mathbb{C}$-linear maps $\psi:\,X\to Y$ with the
intertwining property $\rho_Y(g)\psi = \psi \rho_X(g)$ for all $g \in
G_0\,$. If $X$ is an irreducible $G_0$-representation, then Schur's
lemma says that $\mathrm{Hom}_{G_0}(X,X)$ is one-dimensional, being
spanned by the identity, $\mathrm{Id}_X\,$. For two irreducible
$G_0$-representations $X$ and $Y$, the dimension of $\mathrm
{Hom}_{G_0} (X,Y)$ is either zero or one, by an easy corollary of
Schur's lemma. In the latter case $X$ and $Y$ are said to belong to
the same isomorphism class.

Using the symbol $\lambda$ to denote the isomorphism classes of
irreducible $G_0$-representations, we fix for each $\lambda$ a
standard representation space $R_\lambda\,$. Note that $\mathrm
{dim}\, \mathrm{Hom}_{G_0}(R_\lambda\,, V)$ counts the multiplicity
in $V$ of the irreducible representation of isomorphism class
$\lambda\,$. In our shell-model example with $G_0 = \mathrm{O}_3$ we
have $\lambda = (L,\pi)$, $R_\lambda = \mathbb{C}^{2L+1}$, and
$\mathrm{dim}\, \mathrm{Hom}_{G_0} (R_\lambda\,, V) = m(L,\pi)$.

The following statement can be interpreted as saying that the example
adequately reflects the general situation.
\newtheorem{lem}{Lemma}[section]
\begin{lem}\label{lem:1}
Let $G_0$ act reductively on $V$. Then
\begin{displaymath}
    \bigoplus\nolimits_\lambda \mathrm{Hom}_{G_0}(R_\lambda\, , V)
    \otimes R_\lambda \to V , \qquad \bigoplus\nolimits_\lambda
    (\psi_\lambda \otimes r_\lambda) \mapsto \sum\nolimits_\lambda
    \psi_\lambda(r_\lambda)
\end{displaymath}
is a $G_0$-equivariant isomorphism.
\end{lem}
\noindent{\bf Remark.} The decomposition offered by this lemma
perfectly separates the unitary symmetry multiplets from the
dynamical degrees of freedom and thus gives an immediate view of the
structure of the space of $G_0$-compatible Hamiltonians. Indeed, the
direct sum over isomorphism classes (or `sectors') $\lambda$ is
preserved by the symmetries $G_0$ as well as the compatible
Hamiltonians $H$; and $G_0$ is trivial on $\mathrm{Hom}_{G_0}
(R_\lambda\,,V)$ while the Hamiltonians are trivial on $R_\lambda\,$.

Next, we remove the time-evolution trivial factors $R_\lambda$ from
the picture. To do so, we need to go through the step of transferring
all given structure to the spaces $E_\lambda := \mathrm{Hom}_{G_0}
(R_\lambda\, ,V)$.

\subsubsection{Transfer of structure.}

We first transfer the Hermitian structure of $V$. In the present
setting of a unitary $G_0$-action, the Hermitian scalar product of
$V$ reduces to a Hermitian scalar product on each sector of the
direct-sum decomposition of Lemma \ref{lem:1}, by orthogonality of
the sum. Hence, we may focus attention on a definite sector $E
\otimes R \equiv E_\lambda \otimes R_\lambda\,$. Fixing a
$G_0$-invariant Hermitian scalar product $\langle \cdot , \cdot
\rangle_R$ on $R = R_\lambda$ we define such a product $\langle \cdot
, \cdot \rangle_E : \, E \times E \to \mathbb{C}$ by
\begin{equation}\label{eq:HSE}
    \langle \psi , \psi^\prime \rangle_E := \langle \psi(r) ,
    \psi^\prime(r) \rangle_V / \langle r , r\rangle_R \;,
\end{equation}
which is easily checked to be independent of the choice of $r \in
R\,$, $r \not= 0\,$.

Before carrying on, we note that for any Hermitian vector space $V$
there exists a canonically defined $\mathbb{C}$-antilinear bijection
$C_V : V \to V^\ast$ to the dual vector space $V^\ast$ by $C_V(v) :=
\langle v , \cdot \rangle_V\,$. (In Dirac's language this is the
conversion from `ket' vector to `bra' vector.) By naturalness of the
transfer of Hermitian structure we have the relation $C_{E \otimes R}
= C_E \otimes C_R\,$.

Turning to the more involved step of transferring time reversal $T$,
we begin with a preparation. If $L :\, V \to W$ is a linear mapping
between vector spaces, we denote by $L^t :\, W^\ast \to V^\ast$ the
canonical transpose defined by $(L^t f)(v) = f(Lv)$. Let now $V$ be
our Hilbert space with ket-bra bijection $C \equiv C_V$. Then for any
$g \in \mathrm{U}(V)$ we have the relation $C g C^{-1} = (g^{-1})^t$
because
\begin{equation}
    C (g v) = \langle g v , \cdot \rangle = \langle v , g^{-1} \cdot
    \rangle = (g^{-1})^t C (v) \qquad (v \in V)\;.
\end{equation}
Moreover, recalling the automorphism $G_0 \ni g \mapsto a(g) = T g
T^{-1}$ of $G_0$ we obtain
\begin{equation}
    CT\, g = a(g^{-1})^t\, CT \qquad (g \in G_0)\;.
\end{equation}
Thus, since $C$ and $T$ are bijective, the $\mathbb{C}$-linear
mapping $CT : V \to V^\ast$ is a $G_0$-equivariant isomorphism
interchanging the given $G_0$-representation on $V$ with the
representation on $V^\ast$ by $g \mapsto a(g^{-1})^t$. In particular,
it follows that $T$ stabilizes the decomposition $V = \oplus_\lambda
V_\lambda \simeq \oplus_\lambda (E_\lambda \otimes R_\lambda)$ of
Lemma \ref{lem:1}.

If $T$ exchanges different sectors $V_\lambda\,$, the situation is
very easy to handle (see below). The more challenging case is $T
V_\lambda = V_\lambda\,$, which we now assume.
\begin{lem}\label{lem:2}
Let $T V_\lambda = V_\lambda\,$. Under the isomorphism $V_\lambda
\simeq E_\lambda \otimes R_\lambda$ the time-reversal operator
transfers to a pure tensor
\begin{displaymath}
    T = \alpha \otimes \beta \;, \quad \alpha : \; E_\lambda \to
    E_\lambda \;, \quad \beta : \; R_\lambda \to R_\lambda \;,
\end{displaymath}
with anti-unitary $\alpha$ and $\beta\,$.
\end{lem}
\noindent {\bf Proof.} Writing $E_\lambda \equiv E$ and $R_\lambda
\equiv R$ for short, we consider the transferred mapping $CT : \, E
\otimes R \to E^\ast \otimes R^\ast$, which expands as $CT = \sum
\phi_i \otimes \psi_i$ with $\mathbb{C}$-linear mappings $\phi_i\,,
\psi_i\,$. Since $CT$ is known to be a $G_0$-equivariant isomorphism,
so is every map $\psi_i : \, R \to R^\ast$. By the irreducibility of
$R$ and Schur's lemma, there exists only one such map (up to scalar
multiples). Hence $CT$ is a pure tensor: $CT = \phi \otimes \psi\,$.
Using $C = C_{E \otimes R} = C_E \otimes C_R$ we obtain $T = \alpha
\otimes \beta$ with $\mathbb{C}$-antilinear $\alpha = C_E^{-1} \phi$
and $\beta = C_R^{-1} \psi\,$. Since the tensor product lets you move
scalars between factors, the maps $\alpha$ and $\beta$ are not
uniquely defined. We may use this freedom to make $\beta$
anti-unitary. Because $T$ is anti-unitary, it then follows from the
definition (\ref{eq:HSE}) of the Hermitian structure of $E$ that
$\alpha$ is anti-unitary.

\smallskip\noindent{\bf Remark.} By an elementary argument, which was
spelled out for the anti-unitary operator $T$ in Eq.\
(\ref{eq:Tass}), it follows that $\alpha^2 = \epsilon_\alpha \,
\mathrm{Id}_E$ and $\beta^2 = \epsilon_\beta \, \mathrm{Id}_R$ with
$\epsilon_\alpha\,,\epsilon_\beta \in\{\pm 1\}$. Writing $T^2 =
\epsilon_T \, \mathrm{Id}_V$ we have the relation $\epsilon_\alpha
\epsilon_\beta = \epsilon_T\,$. Thus when $\epsilon_\beta = -1$ the
parity $\epsilon_\alpha = - \epsilon_T$ of the transferred
time-reversal operator $\alpha$ is opposite to that of the original
time reversal $T$.

This change of parity occurs, e.g., in the case of $G_0 = \mathrm
{SU}_2\,$. Indeed, let $R \equiv R_n$ be the irreducible $\mathrm
{SU}_2$-representation of dimension $n+1\,$. It is a standard fact of
representation theory that $R_n$ is $\mathrm{SU}_2$-equivariantly
isomorphic to $R_n^\ast$ by a symmetric isomorphism $\psi = \psi^t$
for even $n$ and skew-symmetric isomorphism $\psi = -\psi^t$ for odd
$n\,$. From $(-1)^n \psi^t = \psi = C_R\, \beta$ and
\begin{equation}\label{eq:psi2beta}
    \psi(v)(v^\prime) = \langle \beta v , v^\prime \rangle_R =
    \overline{\langle \beta^2 v , \beta v^\prime \rangle}_R =
    \langle \beta v^\prime , \beta^2 v \rangle_R = \psi(v^\prime)
    (\beta^2 v)
    \;,
\end{equation}
we conclude that $\beta^2 = (-1)^n \mathrm{Id}_{R_n}\,$.

\subsection{Classification}\label{sect:dyson-classes}

By the decomposition of Lemma \ref{lem:1} the space $Z_{\mathrm{U}
(V)}(G_0)$ of $G_0$-compatible time evolutions in $\mathrm{U}(V)$ is
a direct product of unitary groups,
\begin{equation}
    Z_{\mathrm{U}(V)}(G_0) \simeq
    \prod\nolimits_\lambda \mathrm{U}(E_\lambda) \;.
\end{equation}
We now fix a sector $V_\lambda \simeq E_\lambda \otimes R_\lambda$
and run through the different situations (of which there exist three,
essentially) due to the absence or presence of a transferred
time-reversal symmetry $\alpha : \, E_\lambda \to E_\lambda\,$.

\subsubsection{Class $A$}

The first type of situation occurs when the set $G_1$ of anti-unitary
symmetries is either empty or else maps $V_\lambda \simeq E_\lambda
\otimes R_\lambda$ to a different sector $V_{\lambda^\prime}\,$,
$\lambda \not= \lambda^\prime$. In both cases, the $G$-compatible
time-evolution operators restricted to $V_\lambda$ constitute a
unitary group $\mathrm{U}(E_\lambda) \simeq \mathrm{U}_N$ with $N =
\mathrm{dim}\, E_\lambda$ being the multiplicity of the irreducible
representation $R_\lambda$ in $V$. The unitary groups $\mathrm{U}_N$
or to be precise, their simple parts $\mathrm{SU}_N\,$, are symmetric
spaces (cf.\ Section \ref{sect:sym-spaces}) of the $A$ family or $A$
series in Cartan's notation -- hence the name Class $A$. In random
matrix theory, the Lie group $\mathrm{U}_N$ equipped with Haar
measure is commonly referred to as the Circular Unitary Ensemble,
$\mathrm{CUE}_N\,$ \cite{dyson2}.

The Hamiltonians $H$ in Class $A$ are represented by complex
Hermitian $N \times N$ matrices. By putting a $\mathrm{U}_N
$-invariant Gaussian probability measure
\begin{equation}
    d\mu(H) = c_0\, \mathrm{e}^{- \mathrm{Tr}\,
    H^2 /2\sigma^2} dH \;, \quad dH = \prod_{i=1}^N dH_{ii}
    \prod_{j < k} d H_{jk}\, d H_{kj} \;,
\end{equation}
on that space, one gets what is called the GUE -- the Gaussian
Unitary Ensemble -- defining the Wigner-Dyson \emph{universality
class} of unitary symmetry. An important physical realization of that
class is by electrons in a disordered metal with time-reversal
symmetry broken by the presence of a magnetic field.

\subsubsection{Classes $A$I and $A$II}

We now turn to the cases where $T$ is present and $T V_\lambda =
V_\lambda \simeq E_\lambda \otimes R_\lambda\,$. We abbreviate
$E_\lambda \equiv E$. From Lemma \ref{lem:1} we know that $T = \alpha
\otimes \beta$ is a pure tensor with anti-unitary $\alpha\,$, and we
have $\alpha^2 = \epsilon_\alpha \, \mathrm{Id}_E$ with parity
$\epsilon_\alpha = \epsilon_T \, \epsilon_\beta\,$.

Using conjugation by $\alpha$ to define an automorphism
\begin{equation}
    \tau : \, \mathrm{U}(E) \to \mathrm{U}(E) \;,
    \quad u \mapsto \alpha u \alpha^{-1} ,
\end{equation}
we transfer the conditions (\ref{eq:G-comp}) to $V_\lambda$ and
describe the set $Z_{\mathrm{U}(E)}(G)$ of $G$-compatible time
evolutions in $\mathrm{U}(E)$ as
\begin{equation}
    Z_{\mathrm{U}(E)}(G) =
    \{ x \in \mathrm{U}(E) \mid \tau(x) = x^{-1} \} \;.
\end{equation}
Now let $U \equiv \mathrm{U}(E)$ for short and denote by $K \subset
U$ the subgroup of $\tau$-fixed elements $k = \tau(k) \in U$. The set
$Z_U(G)$ is analytically diffeomorphic to the coset space $U/K$ by
the mapping
\begin{equation}
    U/K \to Z_U(G) \subset U \;, \quad u K \mapsto u \tau(u^{-1}) \;,
\end{equation}
which is called the Cartan embedding of $U/K$ into $U$. The remaining
task is to determine $K$. This is done as follows.

Recalling the definition $C_E \, \alpha = \phi$ and using $C_E \, k =
(k^{-1})^t C_E$ we express the fixed-point condition $k = \tau(k) =
\alpha k \alpha^{-1}$ as $(k^{-1})^t = \phi k \phi^{-1}$ or,
equivalently, $\phi = k^t \phi\, k\,$, which means that the bilinear
form associated with $\phi :\, E \to E^\ast$,
\begin{equation}
    Q_\phi : \; E \times E \to \mathbb{C} \;,
    \quad (e,e^\prime) \mapsto \phi(e)(e^\prime) \;,
\end{equation}
is preserved by $k \in K$. By running the argument around Eq.\
(\ref{eq:psi2beta}) in reverse order (with the obvious substitutions
$\psi\to \phi$ and $\beta\to \alpha$), we see that the non-degenerate
form $Q_\phi$ is symmetric if $\epsilon_\alpha = +1$ and skew if
$\epsilon_\alpha = -1\,$. In the former case it follows that $K =
\mathrm{O}(E) \simeq \mathrm{O}_N$ is an orthogonal group, while in
the latter case, which occurs only if $N \in 2\mathbb{N}\,$, $K =
\mathrm{USp}(E) \simeq \mathrm{USp}_N$ is unitary symplectic. In both
cases the coset space $U / K$ is a symmetric space (cf.\ Section
\ref{sect:sym-spaces}) -- a fact first noticed by Dyson in
\cite{dyson3}.

Thus in the present setting of $T V_\lambda = V_\lambda$ we have the
following dichotomy for the sets of $G$-compatible time evolutions
$Z_{\mathrm{U}(E)}(G) \simeq U/K :$
\begin{displaymath}
\begin{array}{lll}
    \mbox{Class $A$I}: \hspace{0.2cm} &U/K \simeq \mathrm{U}_N
    / \mathrm{O}_N \hspace{0.2cm} &(\epsilon_\alpha = +1) \;,\\
    \mbox{Class $A$II}: &U/K \simeq \mathrm{U}_N /\mathrm{USp}_N
    &(\epsilon_\alpha = -1 \,, \; N \in 2 \mathbb{N}) \;.
\end{array}
\end{displaymath}
Again we are referring to symmetric spaces by the names they -- or
rather their simple parts $\mathrm{SU}_N / \mathrm{SO}_N$ and
$\mathrm{SU}_N / \mathrm{USp}_N$ -- have in the Cartan
classification. In random matrix theory, the symmetric space
$\mathrm{U}_N / \mathrm{O}_N$ (or its Cartan embedding into
$\mathrm{U}_N$ as the symmetric unitary matrices) equipped with
$\mathrm{U}_N$-invariant probability measure is called the Circular
Orthogonal Ensemble, $\mathrm{COE}_N\,$, while the Cartan embedding
of $\mathrm{U}_N / \mathrm{USp}_N$ equipped with $\mathrm{U}_N
$-invariant probability measure is known as the Circular Symplectic
Ensemble, $\mathrm{CSE}_N$ \cite{dyson2}. (Note the confusing fact
that the naming goes by the subgroup which is divided out.)

Examples for Class $A$I are provided by time-reversal invariant
systems with symmetry $G_0 = (\mathrm{SU}_2)_\mathrm{spin}\,$.
Indeed, by the fundamental laws of quantum physics time reversal $T$
squares to $(-1)^{2S}$ times the identity on states with spin $S$.
Such states transform according to the irreducible $\mathrm
{SU}_2$-representation of dimension $2S+1$, and from $\beta^2 =
(-1)^{2S}$ (see the Remark after Lemma \ref{lem:1}) it follows that
$\epsilon_\alpha = \epsilon_T \, \epsilon_\beta = (-1)^{2S} (-1)^{2S}
= +1$ in all cases. A historically important realization of Class
$A$I is furnished by the highly excited states of atomic nuclei as
observed by neutron scattering just above the neutron threshold.

By breaking $\mathrm{SU}_2$-symmetry (i.e., by taking $G_0 = \{
\mathrm{Id} \}$) while maintaining $T$-symmetry for states with
half-integer spin (say single electrons, which carry spin $S = 1/2$),
one gets $\epsilon_\alpha = \epsilon_T = (-1)^{2S} = -1$, thereby
realizing Class $A$II. An experimental observation of this class and
its characteristic wave interference phenomena was first reported in
the early 1980's \cite{bergmann} for disordered metallic magnesium
films with strong spin-orbit scattering caused by gold impurities.

The Hamiltonians $H$, obtained by passing to the tangent space of
$U/K$ at unity, are represented by Hermitian matrices with entries
that are real numbers (Class $A$I) or real quaternions (Class $A$II).
The simplest random matrix models result from putting $K$-invariant
Gaussian probability measures on these spaces; they are called the
Gaussian Orthogonal Ensemble and Gaussian Symplectic Ensemble,
respectively. Their properties delineate the Wigner-Dyson
universality classes of orthogonal and symplectic symmetry.

\sect{Symmetry Classes of Disordered Fermions}\label{sect:10-way}

While Dyson's Threefold Way is fundamental and complete in its
general Hilbert space setting, the early 1990's witnessed the
discovery of various new types of strong universality, which were
begging for an extended scheme:
\begin{itemize}
\item The introduction of QCD-motivated chiral random matrix
    ensembles (reviewed by Verbaarschot in Chapter 32 of this
    handbook) mimicked Dyson's scheme but also transcended it.
\item Number theorists had introduced and studied ensembles of
    $L$-functions akin to the Riemann zeta function (see the
    review by Keating and Snaith in Chapter 24 of this handbook).
    These display random matrix phenomena which are absent in the
    classes $A$, $A$I, or $A$II.
\item The proximity effect due to Andreev reflection, a
    particle-hole conversion process in mesoscopic hybrid systems
    involving metallic as well as superconducting components, was
    found \cite{AZ} to give rise to post-Dyson mechanisms of
    quantum interference (cf.\ Chapter 35 by Beenakker).
\end{itemize}
By the middle of the 1990's, it had become clear that there exists a
unifying mathematical principle governing these post-Dyson random
matrix phenomena. This principle will be explained in the present
section. We mention in passing that a fascinating recent development
\cite{kitaev,ludwig} uses the same principle for a homotopy
classification of topological insulators and
superconductors.

\subsection{Fock space setting}\label{sect:fock}

We now describe an extended setting, which replaces the Hermitian
vector space $V$ by its exterior algebra $\wedge(V)$ but otherwise
retains Dyson's setting to the fullest extent possible. In physics
language we say that we pass from the (single-particle) Hilbert space
$V$ to the fermionic Fock space $\wedge(V)$ generated by $V$. The
$\mathbb{Z}$-grading $\wedge(V) = \oplus_n \wedge^n(V)$ by the degree
$n$ has the physical meaning of particle number. Thus $\wedge^0 (V)
\equiv \mathbb{C}$ is the vacuum, $\wedge^1(V) \equiv V$ is the
one-particle space, $\wedge^2(V)$ is the two-particle space, and so
on. We adhere to the assumption of finite-dimensional $V$. Particle
number $n$ then is in the range $0\leq n\leq N:=\mathrm{dim} \,V$.
Note that $\mathrm{dim} \wedge^n(V) = \begin{pmatrix} N \\ n
\end{pmatrix}$.

The $n$-particle subspace $\wedge^n(V)$ of the Fock space of a
Hermitian vector space $V$ carries an induced Hermitian scalar
product defined by
\begin{equation}
    \big\langle u_1 \wedge \cdots \wedge u_n\, , v_1 \wedge \cdots
    \wedge v_n \big\rangle_{\wedge^n(V)} = \mathrm{Det} \begin{pmatrix}
    \langle u_1 , v_1 \rangle_V &\ldots &\langle u_1 , v_n \rangle_V \\
    \vdots &\ddots &\vdots\\ \langle u_n , v_1 \rangle_V &\ldots
    &\langle u_n , v_n \rangle_V \end{pmatrix} .
\end{equation}
Relevant operations on $\wedge(V)$ are exterior multiplication (or
particle creation) $\varepsilon(v) : \wedge^n(V) \to \wedge^{n+1}(V)$
by $v \in V$ and contraction (or particle annihilation) $\iota(f) :
\wedge^n(V)\to\wedge^{n-1}(V)$ by $f\in V^\ast$. The standard physics
convention is to fix some orthonormal basis $\{ e_k \}_{k = 1,...,N}$
of $V$ and write $a_k^\dagger := \varepsilon(e_k)$ for the particle
creation operators and $a_k := \iota(\langle e_k , \cdot \rangle_V)$
for the particle annihilation operators. This notation reflects the
fact that Hermitian conjugation $\dagger$ in Fock space relates
$\varepsilon(v)$ and $\iota(f)$ by $\varepsilon (v)^\dagger=\iota(f)$
where $f = \langle v,\cdot\rangle_V$. The operators $a_k^\dagger$ and
$a_k$ satisfy the so-called canonical anti-commutation relations
\begin{equation}
    a_k a_l + a_l a_k = 0 = a_k^\dagger a_l^\dagger + a_l^\dagger
    a_k^\dagger \;,\qquad a_k^\dagger a_l^{\vphantom{\dagger}} +
    a_l^{\vphantom{\dagger}} a_k^\dagger = \delta_{kl} \;.
\end{equation}
These represent the defining relations of the Clifford algebra
$\mathrm{Cl}(W)$ of the vector space $W = V \oplus V^\ast$ with
quadratic form $(v \oplus f , v^\prime \oplus f^\prime) \mapsto
f(v^\prime) + f^\prime(v)$.

Having introduced the basic Fock space structure, we now turn to what
is going to be our definition of a symmetry group $G$ acting on Fock
space $\wedge(V)$. As before, we assume that we are given a normal
subgroup $G_0 \subset G$. The action of the elements $g \in G_0$ is
defined by unitary operators on $V$ which are extended to $\wedge(V)$
by
\begin{equation}\label{eq:extend}
    g (v_1 \wedge \cdots \wedge v_n) :=
    (g v_1) \wedge \cdots \wedge (g v_n) \;.
\end{equation}
Similarly, the anti-unitary operator of time reversal $T$ is defined
on $V$ and is extended to $\wedge(V)$ by $T (v_1 \wedge \cdots \wedge
v_n) := (T v_1) \wedge \cdots \wedge (T v_n)$.

Now the $\mathbb{Z}$-grading of Fock space offers the natural option
of introducing another kind of anti-unitary operator, which is a
close cousin of the Hodge star operator for the deRham complex:
particle-hole conjugation, $C$, transforms an $n$-particle state into
an $(N-n)$-particle state or a state with $n$ holes. (Note the change
of meaning of the symbol $C$ as compared to Section
\ref{sect:reduce}.)
\newtheorem{defn}{Definition}[section]
\begin{defn}\label{def:ph-conj}
Fix a generator $\Omega \in \wedge^N(V)$, $N = \mathrm{dim}\, V$,
with normalization $\langle \Omega , \Omega \rangle_{\wedge^N(V)} =
1$. Then particle-hole conjugation $C : \, \wedge^n(V) \to
\wedge^{N-n}(V)$ is the anti-unitary operator defined by
\begin{displaymath}
    (C \psi) \wedge \psi^\prime = \langle \psi , \psi^\prime
    \rangle_{\wedge^n(V)} \, \Omega \;.
\end{displaymath}
\end{defn}
Thus the definition of the operator $C$ uses the Hermitian scalar
product of Fock space and a choice of fully occupied state $\Omega$.
An elementary calculation shows that $C^2 \vert_{\wedge^n(V)} =
(-1)^{n(N-n)}$.

What are the commutation relations of $C$ with $T$ and $g \in G_0\,$?
To answer this question, we observe that by $\mathrm{dim}\, \wedge^N
(V) = 1$ we may always choose $\Omega$ to be $T$-invariant (i.e., $T
\Omega = \Omega$). Then from the following computation,
\begin{align*}
    (TC\psi) \wedge T\psi^\prime &= T \left( (C\psi) \wedge
    \psi^\prime \right) = T \left( \langle \psi , \psi^\prime
    \rangle \, \Omega \right) \cr &= \overline{\langle \psi ,
    \psi^\prime \rangle} \, T \Omega = \langle T \psi , T
    \psi^\prime \rangle \,\Omega = (CT\psi) \wedge T\psi^\prime ,
\end{align*}
we have $CT = TC$. Also, making the natural assumption that both the
vacuum space $\wedge^0(V)$ and the fully occupied space $\wedge^N(V)$
transform as the trivial $G_0$-representation (i.e., $g \Omega =
\Omega$ for $g \in G_0$), a similar calculation gives $C g = g C$.

In order to enlarge the set of possible symmetries and hence the
scope of the theory, we now introduce a `twisted' variant of
particle-hole conjugation. Let $S\in\mathrm{U}(V)$ be an involution
($S^2 = \mathrm{Id}$) and extend $S$ to $\wedge(V)$ by Eq.\
(\ref{eq:extend}). To obtain an extension of the group $G_0\,$, we
require that $S$ commutes with $T$, satisfies $S \Omega = \Omega$,
and normalizes $G_0\,$, i.e., $S G_0 S^{-1} = G_0\,$. (Here we
identify $G_0$ with its action on Fock space.) By twisted
particle-hole conjugation we then mean the operator $\tilde{C} = CS =
SC$. Note that $\tilde{C} G_0 \tilde{C}^{ -1} = G_0$ and $\tilde{C} T
= T \tilde{C}$.
\begin{defn}\label{def:G-Fock}
On the Fock space $\wedge(V)$ over a Hermitian vector space $V$, let
there be the action of a group $G = G_0 \cup T G_0 \cup \tilde{C} G_0
\cup \tilde{C}T G_0$ with $G_0$ a normal subgroup and $\tilde{C} T =
T \tilde{C}$. We call this a {\rm minimal extension} of Dyson's
setting if $G_0$ acts by unitary operators defined on $V$, time
reversal $T$ acts as an anti-unitary operator also defined on $V$,
and twisted particle-hole conjugation $\tilde{C}$ is an anti-unitary
bijection $\wedge^n(V) \to \wedge^{N-n}(V)$.
\end{defn}

\subsection{Classification goal}

The simplest question to ask now is this: what is the structure of
the set of Hamiltonians that operate on Fock space $\wedge(V)$ and
commute with the given $G$-action on $\wedge(V)$? Since this question
ignores the grading of Fock space by particle number, it takes us
back to Dyson's setting and the answer is, in fact, provided by
Dyson's Threefold Way. (Note that in the absence of restrictions, the
most general Hamiltonian in Fock space involves $n$-body interactions
of arbitrary rank $n = 1, 2, 3, \ldots, N$.) So there is nothing new
to discover here.

We shall, however, be guided to a new and interesting answer by
asking a somewhat different question: what is the structure of the
set of \emph{one-body} time evolutions of $\wedge(V)$ which commute
with the given $G$-action? Here by a one-body time evolution we mean
any unitary operator obtained by exponentiating a self-adjoint
Hamiltonian $H$ which is \emph{quadratic} in the particle creation
and annihilation operators:
\begin{equation}\label{eq:H-Fock}
    H= \sum\nolimits_{kl} W_{kl}\,a_k^\dagger a_l^{\vphantom{\dagger}}
    + \frac{1}{2} \sum\nolimits_{k l} \big( Z_{k l}\, a_k^\dagger
    a_l^\dagger + \overline{Z}_{k l}\, a_l a_k \big) \;.
\end{equation}
These operators $U = \mathrm{e}^{-\mathrm{i}tH/\hbar} \in \mathrm{U}
(\wedge V)$ form what is called the spin group, $\mathrm{Spin}(
W_\mathbb{R})$, of the $2N$-dimensional Euclidean $\mathbb{R}$-vector
space $W_\mathbb{R}$ spanned by the Majorana operators $a_k^{
\vphantom{\dagger}} + a_k^\dagger\,$, $\mathrm{i} a_k^{\vphantom{
\dagger}}-\mathrm{i} a_k^\dagger$ ($k = 1, \ldots, N$). $\mathrm
{Spin} (W_\mathbb{R}) \simeq \mathrm{Spin}_{2N}$ is a double covering
of the real orthogonal group $\mathrm{SO}( W_\mathbb{R}) \simeq
\mathrm{SO}_{2N}\,$. The spin group of most prominence in physics is
$\mathrm{Spin}_3 \equiv \mathrm{SU}_2\,$, a double covering of
$\mathrm{SO}_3\,$. (This double covering is known to physicists by
the statement that a rotation by $2\pi$, which acts as the neutral
element of $\mathrm{SO}_3\,$, changes the sign of a spinor.)

Thus, our interest is now in the set
\begin{equation}
    Z_\mathrm{Spin}(G) := Z_{\mathrm{U}(\wedge V)}(G)
    \cap \mathrm{Spin}(W_\mathbb{R})
\end{equation}
of $G$-compatible time evolutions in $\mathrm{Spin}(W_\mathbb{R})
\subset \mathrm{U}(\wedge V)$. By adaptation of the earlier
definition (\ref{eq:G-comp}), the $G$-compatibility conditions are
\begin{equation}\label{eq:G-comp-F}
    U = g_0^{\vphantom{-1}} U g_0^{-1} =
    g_1^{\vphantom{-1}} U^{-1} g_1^{-1}
\end{equation}
for all $g_0 \in G_0 \cup \tilde{C} T G_0$ and $g_1 \in T G_0 \cup
\tilde{C} G_0\,$.

\subsection{Reduction to Nambu space}\label{sect:reduce}

To investigate the set $Z_\mathrm{Spin}(G)$ we use the following
fact. Any invertible element $A \in \mathrm{Cl}(W)$ determines an
automorphism $\gamma \mapsto A \gamma A^{-1}$ of the Clifford algebra
$\mathrm{Cl}(W)$ by conjugation. This conjugation action restricts to
a representation
\begin{equation}
    \tau(g) w := g w g^{-1}
\end{equation}
of $\mathrm{Spin}(W_\mathbb{R}) \subset \mathrm{Cl}(W)$ on $W_\mathbb
{R} \subset \mathrm{Cl}(W)$. Phrased in physics language, the set of
Majorana field operators $w = \sum_k (z_k a_k^\dagger +\overline{z}_k
a_k)\in W_\mathbb{R}$ is closed under conjugation $w\mapsto gwg^{-1}$
by one-body time evolution operators $g\in\mathrm{Spin}(W_\mathbb{R}
)$. In fact, by elementary considerations one finds that $\tau(g): \,
w \mapsto g w g^{-1}$ for $g \in \mathrm{Spin} (W_\mathbb{R})$ is an
orthogonal transformation $\tau(g) \in \mathrm{SO}(W_\mathbb{R})$ of
the Euclidean vector space $W_\mathbb{R} \,$. The mapping $\tau : \,
\mathrm{Spin}(W_\mathbb{R})\to \mathrm{SO}(W_\mathbb{R})$, $g \mapsto
\tau(g)= \tau(-g)$ is two-to-one. It is a covering map, which amounts
to saying that any path in $\mathrm{SO}(W_\mathbb{R})$ lifts uniquely
to a path in $\mathrm{Spin}(W_\mathbb{R})$. Note also that the linear
mapping $\tau(g) : \, W_\mathbb{R} \to W_\mathbb{R}$ extends to a
linear mapping $\tau(g) : \, W \to W$ by $\mathbb{C}$-linearity.

Thus, instead of studying $\mathrm{Spin}(W_\mathbb{R})$ as a group of
operators on the full Fock space $\wedge(V)$, we may simplify our
work by studying its representation $\tau : \, \mathrm{Spin} \to
\mathrm{SO}$ on the smaller space $W = V \oplus V^\ast$, here
referred to as Nambu space. Of course the object of interest is not
$\mathrm {Spin}(W_\mathbb{R})$ but its intersection with the
$G$-compatibility conditions. To keep track of the latter conditions,
we now transfer the $G$-action from $\wedge(V)$ to $W = V \oplus
V^\ast$.

It is immediately clear how to transfer the actions of $G_0$ and $T$,
as these are defined on $V$. In the case of the twisted particle-hole
conjugation operator $\tilde{C}$, we do the following computation.
Let $\psi \in \wedge^n(V)$ and $\psi^\prime \in \wedge^{n+1}(V)$.
Then
\begin{align*}
    (\tilde{C} a_k^\dagger \psi) \wedge \psi^\prime &= \langle
    S a_k^\dagger \psi , \psi^\prime \rangle \, \Omega = \langle
    S \psi , S a_k S^{-1} \psi^\prime \rangle \, \Omega \cr &=
    (\tilde{C} \psi) \wedge (S a_k S^{-1}) \psi^\prime =
    (-1)^{N-n+1} (S a_k S^{-1} \tilde{C} \psi) \wedge \psi^\prime \;.
\end{align*}
Thus the twisted particle-hole conjugate of $a_k^\dagger \in V
\subset \mathrm{Cl}(W)$ is $\tilde{C} a_k^\dagger \tilde{C}^{-1} =
\pm S a_k S^{-1} \in V^\ast \subset \mathrm{Cl}(W)$ where the sign
alternates with particle number. Note that the operation $a_k^\dagger
\mapsto \pm S a_k S^{-1}$ is anti-unitary. Note also that the
untwisted particle-hole conjugation $a_k^\dagger \mapsto a_k^{
\vphantom{\dagger}}$ is none other than the $\mathbb{C}$-antilinear
bijection $C_V : \, V \to V^\ast$, $v \mapsto \langle v , \cdot
\rangle_V\,$.

To sum up, we have the following induced structures on Nambu space:
\begin{itemize}
\item One-body time evolutions $g \in \mathrm{Spin}
    (W_\mathbb{R})$ act on $W = V \oplus V^\ast$ by orthogonal
    transformations $\tau(g) \in \mathrm{SO}(W_\mathbb{R})$.
\item $G_0$ is defined on $V$ and acts on $W = V \oplus V^\ast$
    by $g (v \oplus f) = gv \oplus (g^{-1})^t f$. The same goes
    for time reversal $T$.
\item The operator $\tilde{C}$ of twisted particle-hole
    conjugation induces on $W = V \oplus V^\ast$ an anti-unitary
    involution $V \leftrightarrow V^\ast$.
\end{itemize}
The goal of symmetry classification now is to characterize the set
$Z_\mathrm{SO}(G)$ of elements in $\mathrm{SO}(W_\mathbb{R})$ which
are compatible with the induced $G$-action on $W$. This problem was
posed and solved in \cite{HHZ}, by using an elaboration of the
algebraic tools of Section \ref{sect:dyson} to make a reduction to
the case of the trivial group $G_0 = \{\mathrm{Id}\}$. (The
involution $V \leftrightarrow V^\ast$ given by twisted particle-hole
conjugation is called a \emph{mixing} symmetry in \cite{HHZ}.) The
outcome is as follows.
\newtheorem{thm}{Theorem}[section]
\begin{thm}
The space $Z_\mathrm{SO}(G)$ is a direct product of factors each of
which is a classical irreducible compact symmetric space. Conversely,
every classical irreducible compact symmetric space occurs in this
setting.
\end{thm}
There is no space to reproduce the proof here, but in order to turn
the theorem into an intelligible statement we now record a few basic
facts from the theory of symmetric spaces \cite{helgason,caselle}.

\subsection{Symmetric spaces}\label{sect:sym-spaces}

Let $M$ be a connected $m$-dimensional Riemannian manifold and $p$ a
point of $M$. In some open subset $N_p$ of a neighborhood of $p$
there exists a map $s_p : N_p \to N_p\,$, the geodesic inversion with
respect to $p\,$, which sends a point $x \in N_p$ with normal
coordinates $(x_1, \ldots, x_m)$ to the point with normal coordinates
$(- x_1, \ldots, - x_m)$. The Riemannian manifold $M$ is called
locally symmetric if the geodesic inversion is an isometry (i.e., is
distance-preserving), and is called globally symmetric if $s_p$
extends to an isometry $s_p : M \to M$, for all $p \in M$. A globally
symmetric Riemannian manifold is called a symmetric space for short.

The Riemann curvature tensor of a symmetric space is covariantly
constant, which leads to three distinct cases: the scalar curvature
can be positive, zero, or negative, and the symmetric space is said
to be of compact type, Euclidean type, or non-compact type,
respectively. In random matrix theory each type plays a role: the
first one provides us with the scattering matrices and time
evolutions, the second one with the Hamiltonians, and the third one
with the transfer matrices. Our focus here will be on compact type,
as it is this type that houses the unitary time evolution operators
of quantum mechanics.

Symmetric spaces of compact type arise in the following way. Let $U$
be a connected compact Lie group equipped with a Cartan involution,
i.e., an automorphism $\tau :\, U \to U$ with the involutive property
$\tau^2 = \mathrm{Id}$. Let $K \subset U$ be the subgroup of
$\tau$-fixed points $u = \tau(u)$. Then the coset space $U/K$ is a
compact symmetric space in a geometry defined as follows. Writing
$\mathfrak{u} := \mathrm{Lie} (U)$ and $\mathfrak{k} := \mathrm{Lie}
(K)$ for the Lie algebras of the Lie groups involved, let
$\mathfrak{u} = \mathfrak{k} \oplus \mathfrak{p}$ be the
decomposition into positive and negative eigenspaces of the
involution $d\tau : \, \mathfrak{u} \to \mathfrak{u}$ induced by
linearization of $\tau$ at unity. Fix on $\mathfrak{p}$ a Euclidean
scalar product $\langle \cdot , \cdot \rangle_{\mathfrak{p}}$ which
is invariant under the adjoint $K$-action $\mathrm{Ad}(k) : \,
\mathfrak{p} \to \mathfrak{p}$ by $X \mapsto k X k^{-1}$. Then the
Riemannian metric $g_{uK}$ evaluated on vectors $v, v^\prime$ tangent
to the coset $uK$ is $g_{uK}(v , v^\prime) := \langle dL_u^{-1}(v) ,
dL_u^{-1} (v^\prime) \rangle_{ \mathfrak{p}}$ where $dL_u$ denotes
the differential of the operation of left translation on $U/K$ by $u
\in U$.

It is important for us that the coset space $U/K$ can be realized as
a subset
\begin{equation}\label{eq:Cartan-embed}
    M := \{ x \in U \mid \tau(x) = x^{-1} \}
\end{equation}
by the Cartan embedding $U/K \to M \subset U$, $uK \mapsto u\,
\tau(u^{-1})$. The metric tensor in this realization is given (in a
self-explanatory notation) by $g = \mathrm{Tr}\, dx\, dx^{-1}$. It is
invariant under the $K$-action $M \to M$ by twisted conjugation $x
\mapsto u x \tau(u^{-1})$. The geodesic inversion with respect to $y
\in M$ is $s_y : \, M \to M$, $x \mapsto y x^{-1} y$.

We note that special examples of compact symmetric spaces are
afforded by compact Lie groups $K$. For these examples, one takes $U
= K \times K$ with flip involution $\tau(k,k^\prime) = (k^\prime, k)$
leading to $U/K = (K \times K) / K \simeq K$. Cartan's list of
classical (or large families of) compact symmetric spaces is
presented in Table \ref{table:1}. The form of the generator $H$ of
time evolutions $u = \mathrm{e}^{-\mathrm{i} tH / \hbar}$ is
indicated in the third column, where the notation $W,Z$ refers to the
Fock space expression (\ref{eq:H-Fock}) which translates to $H =
\begin{pmatrix} W &Z \cr Z^\dagger &-W^t \end{pmatrix}$ by the
covering map $\tau :\, \mathrm{Spin} \to \mathrm{SO}\,$.

\begin{table}
\begin{tabular}{|cll|}
\hline\rule[-3mm]{0mm}{8mm}
{\bf family} & {\bf compact type} & {\bf Euclidean type}\\
\hline\rule[-3mm]{0mm}{8mm}
$A$ & $\mathrm{U}_N$ & $H$ complex Hermitian \\ \rule[-3mm]{0mm}{8mm}
$A{\mathrm I}$ & $\mathrm{U}_N/\mathrm{O}_N$ & $H$ real symmetric \\
\rule[-3mm]{0mm}{8mm}
$A\mathrm {II}$ & ${\mathrm U}_{2N}/\mathrm{USp}_{2N}$
& $H$ quaternion self-dual \\ \rule[-3mm]{0mm}{8mm}
$C$ & $\mathrm{USp}_{2N}$ & $Z$ complex symmetric \\
\rule[-3mm]{0mm}{8mm}
$C{\mathrm I}$ &$\mathrm{USp}_{2N}/\mathrm{U}_N$ & $Z$ complex sym.,
$W=0$ \\ \rule[-3mm]{0mm}{8mm}
$B,D$ & $\mathrm{SO}_N$ & $Z$ complex skew \\ \rule[-3mm]{0mm}{8mm}
$D\mathrm{III}$ & $\mathrm{SO}_{2N}/\mathrm{U}_N$ & $Z$ complex skew,
$W=0$ \\ \rule[-3mm]{0mm}{8mm}
$A\mathrm{III}$ &$\mathrm{U}_{p+q}/(\mathrm{U}_p \times \mathrm{U}_q)$
& $Z$ complex $p\times q$, $W=0$ \\ \rule[-3mm]{0mm}{8mm}
$BD{\mathrm I}$ & $\mathrm{O}_{p+q}/(\mathrm{O}_{p}\times
\mathrm{O}_q)$ & $Z$ real $p\times q$, $W=0$ \\ \rule[-3mm]{0mm}{8mm}
$C\mathrm{II}$ & $\mathrm{USp}_{2p+2q} / (\mathrm{USp}_{2p}\times
\mathrm{USp}_{2q})$ & $Z$ quaternion $2p\times 2q$, $W = 0$ \\ \hline
\end{tabular}
\caption{The Cartan table of classical symmetric
spaces}\label{table:1}
\end{table}

\subsection{Post-Dyson classes}

We now run through the symmetry classes beyond those of Wigner-Dyson.
As was mentioned before, these appear in various areas of physics and
in the random matrix theory of $L$-functions. For brevity we
concentrate on their physical realization by quasi-particles in
disordered metals and superconductors.

\subsubsection{Class $D$}

Consider a superconductor with no symmetries in its quasi-particle
dynamics, so $G = \{\mathrm{Id}\}$. (Some concrete physical examples
follow below.) The time evolutions $u = \mathrm{e}^{-\mathrm{i}t H /
\hbar}$ in this case are constrained only by the condition $u \in
\mathrm{Spin}(W_\mathbb{R})$ in Fock space and $\tau(u)\in\mathrm{SO}
(W_\mathbb{R})$ in Nambu space. The orthogonal group $\mathrm{SO}
(W_\mathbb{R}) \simeq \mathrm{SO}_{2N}$ is a symmetric space of the
$D$ family -- hence the name class $D$. In a basis of Majorana
fermions $a_k^{\vphantom{\dagger}} + a_k^\dagger\, $, $\mathrm{i}
a_k^{\vphantom{\dagger}} - \mathrm{i} a_k^\dagger\,$, the matrix of
${\rm i}H \in \mathfrak{so}_{2N}$ is real skew, and that of $H$ is
imaginary skew.

Concrete realizations are found in superconductors where the order
parameter transforms under rotations as a spin triplet in spin space
and as a $p$-wave in real space. A recent candidate for a quasi-2d
(or layered) spin-triplet $p$-wave superconductor is the compound
$\mathrm{Sr}_2 \mathrm{Ru}\, \mathrm{O}_4\,$. (A non-charged analog
is the $A$-phase of superfluid ${}^3 \mathrm{He}$.) Time-reversal
symmetry in such a system may be broken spontaneously, or else can be
broken by an external magnetic field creating vortices in the
superconductor.

The simplest random matrix model for class $D$, the $\mathrm{SO}
$-invariant Gaussian ensemble of imaginary skew matrices, is analyzed
in Mehta's book \cite{mehta}.  From the expressions given there it is
seen that the level correlation functions at high energy coincide
with those of the Wigner-Dyson universality class of unitary symmetry
(Class $A$). The level correlations at low energy, however, show
different behavior defining a separate universality class.  This
universal behavior at low energies has immediate physical relevance,
as it is precisely the low-energy quasi-particles that determine the
thermal transport properties of the superconductor at low
temperatures.

\subsubsection{Class $D$III}

Now, let time reversal $T$ be a symmetry: $G = \{ {\rm Id}, T \}$.
Physically speaking this implies the absence of magnetic fields,
magnetic impurities and other agents which distinguish between the
forward and backward directions of time. Our physical degrees of
freedom are spin-1/2 particles, so $T^2 = - \mathrm{Id}_V\,$.

According to (\ref{eq:G-comp-F}) we are looking for the intersection
$Z_\mathrm{SO}(G)$ of the condition $u^{-1} = T u T^{-1}$ with
$\mathrm{Spin}(W_\mathbb{R})$, or after transfer to Nambu space,
$\mathrm{SO} (W_\mathbb{R})$. By introducing the involution $\tau(u)
:= T u T^{-1}$ we express the wanted set as
\begin{equation}
    Z_\mathrm{SO}(G) = \{u\in \mathrm{SO}(W_\mathbb{R}) \mid u^{-1}
    = \tau(u)\}\;.
\end{equation}
Following the discussion around Eq.\ (\ref{eq:Cartan-embed}) we have
$Z_\mathrm{SO}(G) \simeq U / K$ where $U = \mathrm{SO}(W_\mathbb{R})$
and $K \subset U$ is the subgroup of $\tau$-fixed points in $U$.

In order to identify $K$ we note that time reversal $T : \, W \to W$
preserves the real subspace $W_\mathbb{R}$ of Majorana operators
$a_k^{\vphantom{\dagger}} + a_k^\dagger\,$, $\mathrm{i} a_k^{
\vphantom{\dagger}} - \mathrm{i} a_k^\dagger\,$. Because $T^2 = -
\mathrm{Id}$, the $\mathbb{R}$-linear operator $T : \, W_\mathbb{R}
\to W_\mathbb{R}$ is a complex structure of the real vector space
$W_\mathbb{R} \simeq \mathbb{R}^{2N} \simeq \mathbb{C}^N $. In other
words, there exists a basis $\{ e_{1,\,j}\, , e_{2,\,j} \}_{j = 1,
\ldots, N}$ of $W_\mathbb{R}$ such that $T e_{1,\,j} = e_{2,\,j}$ and
$T e_{2,\,j} = - e_{1,\,j}\,$. Now the $\tau$-fixed point condition
$k = \tau(k)$ says that $k \in K$ commutes with the complex linear
extension $J : W \to W$ of $T : W_\mathbb{R} \to W_\mathbb{R}$ by $J
e_{\pm , \, j} = \pm \mathrm{i} e_{\pm , \, j}$ where $e_{\pm , \, j}
= e_{1,\,j} \pm \mathrm{i} e_{2,\,j}\,$. The general element $k$ with
this property is a $\mathrm{U}_N $-transformation which acts on
$\mathrm{span}_\mathbb{C} \{ e_{+, \,1}, \ldots, e_{+,\,N} \}$ as $k$
and on $\mathrm{span}_\mathbb{C} \{ e_{-,\,1}, \ldots, e_{-,\,N} \}$
as $\overline{k} = (k^{-1})^t$. Hence $K = \mathrm{U}_N$ and
\begin{equation}
    Z_\mathrm{SO}(G)\simeq U/K \simeq \mathrm{SO}_{2N}/\mathrm{U}_N\;,
\end{equation}
which is a symmetric space in the $D$III family.

Known realizations of this symmetry class exist in gapless
superconductors, say with spin-singlet pairing, but with a sufficient
concentration of spin-orbit impurities to break spin-rotation
symmetry. In order for quasi-particle excitations to exist at low
energy, the spatial symmetry of the order parameter should be
different from $s$-wave. A non-charged realization occurs in the
$B$-phase of ${}^3 \mathrm{He}$, where the order parameter is
spin-triplet without breaking time-reversal symmetry. Another
candidate are heavy-fermion superconductors, where spin-orbit
scattering often happens to be strong owing to the presence of
elements with large atomic weights such as uranium and cerium.

\subsubsection{Class $C$}

Next let the spin of the quasi-particles be conserved, but let
time-reversal symmetry be broken instead. Thus magnetic fields (or
some equivalent $T$-breaking agent) are now present, while the effect
of spin-orbit scattering is absent. The symmetry group of the
physical system then is $G = G_0 = \mathrm{Spin}_3 = \mathrm {SU}_2\,
$. Such a situation is realized in spin-singlet superconductors in
the vortex phase. Prominent examples are the cuprate superconductors,
which are layered and exhibit an order parameter with $d$-wave
symmetry in their copper-oxide planes.

The symmetry-compatible time evolutions are identified by going
through the process of eliminating the unitary symmetries $G_0 = G$.
For that, we decompose the Hilbert space as $V = E \otimes R\,$, $E =
\mathrm{Hom}_G(R,V)$, where $R := \mathbb{C}^2$ is the fundamental
representation of $G = \mathrm{SU}_2\,$. Now there exists a
skew-symmetric $\mathrm{SU}_2$-equivariant isomorphism (known in
physics by the name of spin-singlet pairing) between the vector space
$R$ and its dual $R^\ast$. Therefore we have $W = V \oplus V^\ast
\simeq (E \oplus E^\ast) \otimes R\,$, and elimination of the
conserved factor $R$ transfers the canonical symmetric form of $W = V
\oplus V^\ast$ to the canonical alternating form $(e \oplus f ,
e^\prime \oplus f^\prime) \mapsto f(e^\prime) - f^\prime(e)$ of $E
\oplus E^\ast$. On transferring also the Hermitian scalar product
from $V \oplus V^\ast$ to $E \oplus E^\ast$, one sees that the
$G$-compatible time evolutions form a unitary symplectic group,
\begin{equation}
    Z_\mathrm{SO}(G) = \mathrm{SO}(W_\mathbb{R})^G \simeq
    \mathrm{USp}(E \oplus E^\ast) \;,
\end{equation}
which is a compact symmetric space of the $C$ family.

\subsubsection{Class $C$I}\label{sec:classCI}

The next class is obtained by taking spin rotations as well as the
time reversal $T$ to be symmetries of the quasi-particle system. Thus
the symmetry group now is $G = G_0 \cup T G_0$ with $G_0 =
\mathrm{Spin}_3 = \mathrm{SU}_2\,$. Like in the previous symmetry
class, physical realizations are provided by the low-energy
quasi-particles of unconventional spin-singlet superconductors. The
superconductor must now be in the Meissner phase where magnetic
fields are expelled by screening currents.

To identify the relevant symmetric space, we again transfer from $V
\oplus V^\ast = (E \oplus E^\ast) \otimes R$ to the reduced space $E
\oplus E^\ast$. By this reduction, the canonical form undergoes a
change of type from symmetric to alternating as before. We must also
transfer time reversal; because the fundamental representation $R =
\mathbb{C}^2$ of $\mathrm{SU}_2$ is self-dual by a skew-symmetric
isomorphism, the parity of the time-reversal operator changes from
$T^2 = - \mathrm{Id}_{V \oplus V^\ast}$ to $T^2 = + \mathrm{Id} _{E
\oplus E^\ast}$ by the mechanism explained at the end of Section
\ref{sect:dyson-classes}.

We have $Z_\mathrm{SO}(G) \simeq U/K$ where $U = \mathrm{USp}(E
\oplus E^\ast)$ and $K$ is the subgroup of elements fixed by
conjugation with $T$. Because the reduced $T$ squares to $+1$, we may
realize it on matrices as the complex conjugation operator by working
in a basis of $T$-fixed vectors of $E \oplus E^\ast$. The Lie algebra
elements $X \in \mathfrak{usp}(E \oplus E^\ast)$ have the form $X =
\begin{pmatrix} A &B \cr -\overline{B} &\overline{A}
\end{pmatrix}$ with anti-Hermitian $A$ and complex symmetric $B$.
They commute with the operation of complex conjugation if $A$ is real
skew and $B$ real symmetric. Matrices $X$ with such $A$ and $B$ span
the Lie algebra $\mathfrak{u}_N$, $N = \mathrm{dim}(E)$. At the Lie
group level it follows that $K = \mathrm{U}_N\,$. Hence $Z_{SO}(G)
\simeq \mathrm{USp}_{2N} / \mathrm{U}_N$ -- a symmetric space in the
$C$I family.

\subsubsection{Class $A$III}

So far, we have made no use of twisted particle-hole conjugation
$\tilde{C}$ as a symmetry, but now let the symmetry group be $G = G_0
\cup \tilde{C} G_0$ where $G_0 = \mathrm{U}_1$ acts on $W = V \oplus
V^\ast$ by $v \oplus f \mapsto zv \oplus z^{-1} f$ (for $z \in
\mathbb{C}\,$, $|z|=1$).

In order for the elements $u \in \mathrm{SO}(W_\mathbb{R})$ to
commute with the $G_0$-action, they must be of the block-diagonal
form $u = k \oplus (k^{-1})^t$, $k \in \mathrm{U}(V)$. Therefore
$Z_\mathrm{SO}(G_0) \simeq \mathrm{U}(V)$. The wanted set then is
$Z_\mathrm{SO}(G) \simeq U / K$ with $U \equiv \mathrm{U}(V)$ and $K$
the subgroup of elements which are fixed by conjugation with
$\tilde{C}$.

Recall from Section \ref{sect:fock} that $\tilde{C}\vert_{V} = C S$
where $S \in U$, $S^2 = \mathrm{Id}$, and untwisted particle-hole
conjugation $C$ coincides (up to an irrelevant sign) with the
canonical bijection $C_V : \, V \to V^\ast$, $C_V(v) = \langle v ,
\cdot \rangle_V$. The condition for $u = k \oplus (k^{-1})^t$ to
belong to $K$ reads
\begin{equation}
    (k^{-1})^t = \tilde{C} k \,\tilde{C}^{-1} .
\end{equation}
Since $k^{-1} = k^\dagger$ and $C^{-1} k^t C = k^\dagger$, this
condition is equivalent to $k = S k S\,$.

Now let $V = V_+ \oplus V_-$ where $V_\pm$ are orthogonal subspaces
with projection operators $\Pi_\pm\,$. Then if $S = \Pi_+ - \Pi_-$ we
have $K = \mathrm{U}(V_+) \times \mathrm{U}(V_-)$ and hence
\begin{equation}\label{eq:AIII}
    Z_\mathrm{SO}(G) \simeq \mathrm{U}(V) / (\mathrm{U}(V_+) \times
    \mathrm{U}(V_-))
\end{equation}
or $Z_\mathrm{SO}(G) \simeq \mathrm{U}_N / (\mathrm{U}_p \times
\mathrm{U}_{N-p})$ with $p = \mathrm{dim}\, V_+\,$.

The space (\ref{eq:AIII}) is a symmetric space of the $A$III family.
Its symmetry class is commonly associated with random-matrix models
for the low-energy Dirac spectrum of quantum chromodynamics with
massless quarks \cite{Ver94}. An alternative realization exists
\cite{asz} in $T$-invariant spin-singlet superconductors with
$d$-wave pairing and soft impurity scattering.

\subsubsection{Classes $BD$I and $C$II}

Finally, let the symmetry group $G$ have the full form of Definition
\ref{def:G-Fock}, with $G_0 = \mathrm{U}_1$ and $\tilde{C}$ as before
(Class $A$III) and a time-reversal symmetry $T$, $T^2 = \pm
\mathrm{Id}\,$. We recall that the elements of $Z_\mathrm{SO}(G_0)$
are $u = k \oplus (k^{-1})^t$, $k \in \mathrm{U} (V)$. The
requirement of commutation with the product $\phi := \tilde{C}T : \,
V \to V^\ast$ of anti-unitary symmetries is equivalent to the
condition $\phi = k^t \phi\, k$.

Let $U := Z_\mathrm{SO}(G_0 \cup \tilde{C}T G_0)$. To identify $U$,
we use that $\phi(v)(v^\prime) = \langle ST v , v^\prime \rangle$ and
$ST = TS$. By the computation of (\ref{eq:psi2beta}), it follows that
the parity of $T$ equals the parity of the isomorphism $\phi : \, V
\to V^\ast$. In other words, if $T^2 = \epsilon\, \mathrm{Id}$ then
$\phi^t = \epsilon\, \phi\,$. Thus the condition $\phi = k^t \phi\,k$
singles out an orthogonal group $U=\mathrm{O}(V) \simeq \mathrm{O}_N$
in the symmetric case ($\epsilon = +1$) and a unitary symplectic
group $U = \mathrm{USp}(V) \simeq \mathrm{USp}_N$ in the alternating
case ($\epsilon = -1$).

In both cases, the wanted set is $Z_\mathrm{SO}(G) = U/K$ with $K$
the subgroup of fixed points $k = \tilde{C}^{-1} (k^{-1})^t \tilde{C}
= S k S$. In the former case we have $K \simeq \mathrm{O}_p \times
\mathrm{O}_{N-p}\,$, and in the latter case $K \simeq \mathrm{USp}_p
\times \mathrm{USp}_{N-p}$ (with even $N$, $p$). Thus we arrive at
the final two entries of Cartan's list:
\begin{displaymath}
\begin{array}{lll}
    \mbox{Class $BD$I}: &U/K \simeq \mathrm{O}_N / (\mathrm{O}_p
    \times \mathrm{O}_{N-p}) &(T^2 = +1)\, , \\ \mbox{Class $C$II}:
    &U/K \simeq \mathrm{USp}_N / (\mathrm{USp}_p \times
    \mathrm{USp}_{N-p}) &(T^2 = -1)\, .
\end{array}
\end{displaymath}
These occur as symmetry classes in the context of the massless Dirac
operator \cite{Ver94}. Class $BD$I is realized by taking the gauge
group to be $\mathrm{SU}_2$ or $\mathrm{USp}_{2n}\,$, Class $C$II by
taking fermions in the adjoint representation or gauge group
$\mathrm{SO}_n \,$.

\sect{Discussion}

Given the classification scheme for disordered fermions, it is
natural to ask whether an analogous scheme can be developed for the
case of bosons. Although there exists no published account of it
(see, however, \cite{LSZ06}), we now briefly outline the answer to
this question.

The mathematical model for the bosonic Fock space is a symmetric
algebra $\mathrm{S}(V)$. It is still equipped with a canonical
Hermitian structure induced by that of $V$. The real form
$W_\mathbb{R}$ of Nambu space $W = V \oplus V^\ast$ for bosons has an
interpretation as a classical phase space spanned by positions $q_j =
(a_j^{\vphantom{\dagger}} + a_j^\dagger)/\sqrt{2}$ and momenta $p_j =
(a_j^{\vphantom{\dagger}} - a_j^\dagger) /\sqrt{2}\,\mathrm{i}\,$. At
the level of one-body unitary time evolutions in Fock space, the role
of the spin group $\mathrm{Spin} (W_\mathbb{R})$ for fermions is
handed over to the metaplectic group $\mathrm{Mp}(W_\mathbb{R})$ for
bosons.

By the quantum-classical correspondence, a one-parameter group of
time evolutions $u_t = \mathrm{e}^{- \mathrm{i} t H / \hbar} \in
\mathrm{Mp}(W_\mathbb{R})$ in Fock space gets assigned to a linear
symplectic flow $\tau(u_t) \in \mathrm{Sp} (W_\mathbb{R})$ in
classical phase space. This correspondence $\tau : \, \mathrm{Mp}
(W_\mathbb{R}) \to \mathrm{Sp} (W_\mathbb{R})$ is still two-to-one
(reflecting, e.g., the well-known fact that the sign of the harmonic
oscillator wave function is reversed by time evolution over one
period). An important difference as compared to fermions is that the
classical flow $\tau (u_t) \in \mathrm{Sp} (W_\mathbb{R})$ is not
unitary in any natural sense.

In nuclear physics, the differential equation of the flow $\tau(u_t)$
is called the RPA equation. For example, in the case without
symmetries this equation reads
\begin{equation}
    \frac{d}{dt}\,a_k^\dagger = \sum\nolimits_j \big( a_j^\dagger
    A_{jk} + a_j B_{jk} \big)\,, \quad \frac{d}{dt}\,a_k =
    \sum\nolimits_j \big( a_j^\dagger C_{jk} + a_j D_{jk} \big)\,,
\end{equation}
where one requires $B = B^t$, $C = C^t$, and $D = -A^t$ in order for
the canonical commutation relations of the boson operators
$a^\dagger, a$ to be conserved. Unitarity of the flow (as a time
evolution in Fock space) requires $A = - A^\dagger$ and $C =
B^\dagger$. This should be compared with the fermion problem in Class
$C$, where one has exactly the same set of equations but for a single
sign change: $C = - B^\dagger$. Thus the corresponding generator of
time evolution is $X = \begin{pmatrix} A &B \cr \pm\overline{B}
&\overline{A} \end{pmatrix}$ where the plus sign applies to bosons
and the minus sign to fermions. In either case $X$ belongs to the
\emph{same} complex Lie algebra, $\mathfrak{sp}(W)$. The difference
is that the generator for fermions lies in a \emph{compact} real form
$\mathfrak{usp}(W) \subset \mathfrak{sp}(W)$, whereas the generator
for bosons lies in a \emph{non-compact} real form $\mathfrak{sp}
(W_\mathbb{R}) \subset \mathfrak{sp}(W)$.

This remains true in the general case with symmetries. Thus if the
word `symmetry class' is understood in the complex sense, then the
bosonic setting does not lead to any new symmetry classes; it just
leads to different real forms of the known symmetry classes viewed as
complex spaces. The same statement applies to the non-Hermitian
situation. Indeed, all of the spaces of \cite{magnea} are complex or
non-compact real forms of the symmetric spaces of Cartan's table.
Here we must reiterate that the notion of symmetry class is an
algebraic one whose prime purpose is to inject an organizational
principle into the multitude of possibilities. It must not be
misunderstood as a cheap vehicle to produce immediate predictions of
eigenvalue distributions and universal behavior!

Let us end with a few historical remarks. The disordered harmonic
chain, a model in the post-Dyson Class $BD$I, was first studied by
Dyson \cite{dyson:chain}. The systematic field-theoretic study of
models with sublattice symmetry (later recognized as members of the
chiral classes $A$III, $BD$I) was initiated by Oppermann, Wegner and
Gade \cite{OppWeg,gade,GadeWeg}. Gapless superconductors were the
subject of numerous papers by Oppermann; e.g., \cite{oppermann}
computes the one-loop beta function of the non-linear sigma model for
Class $C$I.

The 10-way classification of Section \ref{sect:10-way} was originally
discovered by a very different reasoning: the mapping of random
matrix problems to effective field-theory models \cite{suprev}
combined with the fact that closure of the renormalization group flow
takes place for non-linear sigma models where the target is a
symmetric space. A less technical early confirmation of the 10-way
classification came from Wegner's flow equations \cite{wegner-flow}.
These take the form of a double-commutator flow for Hamiltonians $H$
belonging to a matrix space $\mathfrak{p}$; if the double commutator
$[\mathfrak{p} , [\mathfrak{p} , \mathfrak{p}]]$ closes in
$\mathfrak{p}$, so does Wegner's flow. The closure condition is
satisfied precisely if $\mathfrak{p}$ is the odd part of a Lie
algebra $\mathfrak{u} = \mathfrak{k} \oplus \mathfrak{p}$ with
involution, i.e., the infinitesimal model of a symmetric space.

Last but not least, let us mention the viewpoint of Volovik (see,
e.g., \cite{volovik}) who advocates classifying single-particle
Green's functions rather than Hamiltonians. That viewpoint in fact
has the advantage that it is not tied to non-interacting systems but
offers a natural framework in which to include (weak) interactions.

\end{document}